\documentclass[11pt,A4paper,twoside,openright]{article}
\usepackage{graphicx,amssymb,epsfig}
\author{Ji\v r\'i Bi\v c\'ak$^{1}$\footnote{e-mail: bicak@mbox.troja.mff.cuni.cz}
        \ and David Kubiz\v n\'ak$^{1,2}$\footnote{e-mail: kubiznak@phys.ualberta.ca}\\\\
{\it $^1$Institute of Theoretical Physics, Charles University in Prague,}\\
{\it V Hole\v sovi\v ck\'ach 2, 180 00 Prague 8, Czech Republic}\\\\
{\it $^2$Theoretical Physics Institute, University of Alberta,}\\
{\it  412 Physics Building, Edmonton, AB, Canada, T6G 2J1 }\\
}
\title{\bf Ultrarelativistic sources in nonlinear electrodynamics}
\begin{document}
\maketitle
\bibliographystyle{plain}
\thispagestyle{plain}
\begin{abstract}
      {\footnotesize
        The fields of rapidly moving sources are studied within nonlinear electrodynamics by boosting the fields of
        sources at rest.  As a consequence of the ultrarelativistic limit the
        $\delta$--like electromagnetic shock waves are found. The character of the field within the shock depends on the
         theory of nonlinear electrodynamics considered. In particular, we obtain the field of an ultrarelativistic
         charge in the Born--Infeld theory.}
\end{abstract}
\section{Introduction}
It was realized a long time ago within Maxwell's theory that the
electromagnetic field of a uniformly moving charged particle with
velocity $\beta$ approaching velocity of light $(c=1)$ is
approximately the same as the field of a pulse of a plane wave.
This similarity was exploited in the studies of the
electromagnetic interactions of relativistic particles by Fermi,
Weizs\H acker, Williams and others---see, e.g., \cite{Jackson} for
a brief review. A rigorous treatment of the limiting field arising
when $\beta=1$ was studied much later---after the spacetime
representing a Schwarzschild black hole boosted to the velocity of
light was found \cite{Aich}, and ultrarelativistic limits proved
to be of a great value also in general relativity and other
gravity theories. The limiting fields have been used as `incoming
states' in the scattering processes with high initial speeds,
including the quantum scattering of two pointlike particles at
center--of--mass energies higher or equal to the Planck energy.
This quantum process has been shown to have close connection with
classical black hole collisions at the speed of light---see, e.g.,
\cite{Hooft}, \cite{San}, \cite{Fabb}, \cite{Death}.

A careful study of the ultrarelativistic limits---or so--called
`lightlike contractions'---of electromagnetic fields in Minkowski
space within Maxwell's theory was done in 1984 by Robinson and
R\'ozga \cite{Rob}. They considered lightlike contractions of
general multipole fields and have established, as expected, that
the field gets concentrated on null hyperplanes where it shows the
$\delta$--like profile, and resembles a plane wave. The multipole
structure of higher order than monopole can be preserved in such a
limit. All these fields possess pointlike singularities travelling
with the speed of light. They survive the lightlike contractions.
A `lightlike singularity' is present also in the Aichelburg--Sexl
metric arising from the lightlike contraction of a family of
Schwarzschild spacetimes \cite{Aich, Emb}.

In this paper we study the ultrarelativistic limits in the
theories of nonlinear electrodynamics (NLE) which represent models
of classical singularity--free theories with acceptable concepts
of point charges (see, e.g., \cite{Bial}, \cite{Plebanski} for a
review). Nonlinear electromagnetic effects in vacuum are under
investigation by experimentalists (e.g. \cite{Fron}), however,
Maxwell's theory and quantum electrodynamics are in a remarkably
good agreement with an experiment until now. Our motivation stems
from a basic question whether, and if yes then how, finiteness of
the fields of static charges in a suitable NLE survives the
ultrarelativistic limit, and also from the interest in NLE, most
notably in the Born-Infeld (BI) theory \cite{BI}, which arose
relatively recently owing to the developments in string theories
(see, e.g., \cite{Zwieb}). In these theories the BI theory appears
as an effective theory at different levels \cite{Tsie}, especially
in  connection with $p$--branes \cite{Thor}. For example the
motion of a single isolated $(p+1)$--dimensional D--brane moving
in a flat $(d+1)$--dimensional spacetime is governed by so--called
Dirac--Born--Infeld action
\begin{equation}\label{brany}
  S_{\rm DBI}=\int\!d^{p+1}x\sqrt{-{\rm det}(G_{\mu\nu}+2\pi\alpha'F_{\mu\nu})},\quad G_{\mu\nu}
  =\eta_{mn}\frac{\partial z^{m}}{\partial x^{\mu}}\frac{\partial z^{n}}{\partial x^{\nu}},
\end{equation}
where $m,n=0,...d,\ \mu,\nu=0,...p$, $G_{\mu\nu}$ is Minkowskian
metric induced on D--brane,
$F_{\mu\nu}=\partial_{\mu}A_{\nu}-\partial_{\nu}A_{\mu}$, and
$\alpha'$ is the inverse string tension. In {`Monge gauge'} the
vector field $A_{\mu}$ directly satisfies the Born--Infeld field
equations in $(p+1)$--dimensional spacetime \cite{Gib}.

Even at classical level the position of the BI theory among other
NLEs is exceptional. As some other NLEs it yields finite static
fields with finite energy. However, it is the {\em only} NLE in
which the speed of light does not depend on the polarization, i.e.
no birefringence phenomena occurs \cite{Plebanski, Boil}. The BI
system of equations can be enlarged as a system of hyperbolic
conservation laws with remarkable properties which recently
attracted attention (see \cite{Brenier} and references therein).

Our paper is organized as follows: In the next section the concept
of ultrarelativistic limit is revised. In the third section a
spherically symmetric solution of NLE is summarized. The main
results are formulated in the Section 4 where the procedure how to
find the ultrarelativistic fields of a charge coupled to an
arbitrary NLE is presented and consequently applied on the
Born--Infeld charge (BIon \cite{Gib}). Finally follow the
concluding remarks.

\section{The ultrarelativistic limit}
Following closely \cite{Rob}, \cite{Emb}, we first briefly
describe what we mean by the ultrarelativistic limit of the
electromagnetic fields in mathematical terms. Denote by $(M,{\bf
g})$ Minkowski space, with a chart $X: M\,\rightarrow\, \mathbb{
R}^4$ with coordinates $x^{\mu}$ in which
$g_{\mu\nu}=\eta_{\mu\nu}$ is a Lorentzian metric. Next, consider
Lorentz boosts $\Phi_{\beta}:M\,\rightarrow\, M,\;
\vert{\beta}\vert<1$, characterized by charts $X_{(\beta)}: M\
\rightarrow\ {\bf R}^4$ with coordinates $x_{(\beta)}^{\mu}$,
i.e., $x^{\mu}=\Lambda^{\ \ \ \mu}_{(\beta)\
\;\nu}x^{\nu}_{(\beta)}$. For an arbitrary tensor field ${\bf T}$
on $M$, the motions $\Phi_{\beta}$ induce one--parameter family of
the fields $\star\Phi_{\beta}{\bf T}$ on $M$, where $\star\Phi$ is
`derivative of $\Phi$' (see e.g. \cite{Wald}, Appendix C). If an
electromagnetic field tensor ${\bf F}$ is given on $M$,
$\star\Phi_{\beta}{\bf F}$ denotes the corresponding family of
`boosted' electromagnetic fields; we also have
$\star\Phi_{\beta}{\bf g}={\bf g}$. We understand the
ultrarelativistic---or lightlike---limit as a distributional limit
$\vert\beta\vert\to 1$ of this family.

In the case of linear theories one can prove, using the theory of
distributions, that the ultrarelativistic limit `commutes with
field equations', i.e., the fields after the limit satisfy again
the field equations. With nonlinear theories one could turn to
Colombeau theory \cite{Coul}, which provides a possible framework
for studying nonlinear operations with singular functions.
However, we do not attempt this here---in fact, we do not need to
do it. In the following we shall see that in case of nonlinear
electrodynamics the fields obtained by the ultrarelativistic limit
are  well behaved functions within distribution theory.

Let us conclude this section by quoting two mathematical lemmas which will
be needed in Section 4 (for proofs see \cite{Gwaiz}, \cite{Antos}):\\
{\bf Lemma 1.} {\it Let $h\geq 0$ is a Lebesgue integrable function in
$\mathbb{R}$ and  $\int\!h(x)dx=1.$ Then $\lim\limits_{A\to
0}\frac{1}{A}
h\!\left(\frac{x}{A}\right)\stackrel{D'}{=}\delta(x)$,
where $D'$ denotes the limit in a distributional sense.}\\
{\bf Lemma 2.} {\it Let $\alpha_{\lambda}$ is a sequence of smooth ($
C^{\infty}$) functions on a compact region $\Omega\subset \mathbb{
R}^{n}$, which converge uniformly to a smooth function $\alpha$,
and the same is true for all their derivatives. Let a sequence of
distributions $g_{\lambda}$ converges to a distribution $g$,
$g_{\lambda}\!\stackrel{D'(\Omega)}{\longrightarrow} g$. Then
$\alpha_{\lambda}g_{\lambda}\stackrel{D'(\Omega)}{\,\longrightarrow\,}\alpha
g$. If furthermore
$\left(\partial_{1}\alpha_{(\lambda)}(x)\right)^{2}+\dots+\left(\partial_{n}
\alpha_{(\lambda)}(x)\right)^{2}\nolinebreak\neq 0$, then
$g_{\lambda}(\alpha_{\lambda})\stackrel{D'(\Omega)}{\longrightarrow}
g(\alpha).$}\\
\section{Spherically symmetric solutions}
Let $F_{\mu\nu}$ is the electromagnetic field tensor,
$F_{\mu\nu}=\partial_{\mu}A_{\nu}-\partial_{\nu}A_{\mu}$. The
theories of nonlinear electrodynamics start out from the
Lagrangian density $\pounds$, which is an arbitrary function of
the invariant $F=\frac{1}{4}\,F_{\mu\nu}F^{\mu\nu}$ and of the
square of the pseudoinvariant
$G=\frac{1}{4}\,F_{\mu\nu}{\,}^{\star}\!{F^{\mu\nu}}$,
$\pounds=\pounds(F,G^{2}).$ Since a nonlinear behavior may prove
significant only in very strong fields, one commonly requires the
{principle of correspondence} (POC): in a weak--field limit NLEs
have to approach the linear Maxwell theory that is so well
experimentally confirmed. The action of NLE describing charge $Q$
interacting with electromagnetic field reads
\begin{equation}\label{S}
S=-Q\!\int\! A_{\mu}(z)
dz^{\mu}+\frac{1}{4\pi}\int\! d^{4}x\sqrt{-g}\pounds(F,G^{2}),
\end{equation}
where, following POC, we impose the condition
\begin{equation}\label{POC}
\quad
\lim\limits_{F^{\mu\nu}\to\, 0} \pounds=F+O(F^{2},G^{2}).
\end{equation}
The resulting field equations have the form
\begin{equation}\label{f}
\left(\pounds_{,F}F^{\mu\nu}+\pounds_{,G}{\,}^{\star}\!{F^{\mu\nu}} \right)_{;\,\nu}
=4\pi\frac{Q}{\sqrt{-g}}\int\!dz^{\mu}\delta(x-z).
\end{equation}
Equations (\ref{S}) and (\ref{f}) are written in a general
covariant form, $g={\rm det}\, g_{\mu\nu}$, semicolon denotes
covariant derivative, $\pounds_{,F}=\partial \pounds/\partial F$,
etc.

One can show that spherically symmetric solutions of (\ref{f})
must necessarily be static \cite{Slavik}. In spherical coordinates
$(t,r,\theta,\varphi)$ one finds
\begin{equation} \label{Fss}
  \mbox{{\bf F}}=-E dt\land dr,
\end{equation}
where the radial component of the electric field $E(r)$ is
implicitly given by the relation
\begin{equation}\label{Er}
  \pounds_{,F}(F,G)E=\frac{Q}{r^{2}},\quad F=-\frac{1}{2}\,E^2,\quad G=0.
\end{equation}
In particular, for the Born--Infeld theory the Lagrangian density
$\pounds_{\rm BI}$ and the spherically symmetric solution
(\ref{Fss}) read
\begin{equation}\label{BI}
\pounds_{\rm BI}=
b^{2}\left(\sqrt{1+\frac{2F}{b^2}-\frac{G^{2}}{b^{4}}}-1\right),\qquad
E_{\rm BI}=\frac{Q}{\sqrt{r^{4}+r_{0}^{4}}} \ \mbox{,}\quad
r_{0}=\sqrt{\frac{Q}{b}};
\end{equation}
here parameter $b$ plays the role of a limiting field value. The
Maxwell theory is recovered in the weak field limit, $b\to
\infty$. The spherically symmetric solution (\ref{BI}) is finite
everywhere, though at the origin the electric field ceases to be
smooth. It yields a finite energy.

\section{Ultrarelativistic charges in nonlinear electrodynamics}
We first transform the spherically symmetric solutions (\ref{Fss})
into cylindrical coordinates $(t,z,\rho,\varphi)$ and then apply
the Lorentz boost along the negative $z$--axis,
\begin{equation}\label{Boost}
  z=\gamma(Z+\beta T),\quad t=\gamma(T+\beta Z),\qquad\gamma=\frac{1}{\sqrt{1-\beta^2}}.
\end{equation}
The resulting field in the boosted frame $(T, Z, \rho, \varphi)$
turns out to have the form
\begin{equation}\label{Fbost}
 \star\Phi_{\beta}\mbox{{\bf F}}=-E^{Z}dT\land dZ-E^{\rho}dT\land d\rho-B^{\varphi} dZ\land d\rho,
\end{equation}
with components
\begin{equation}\label{boosted}
E^{Z}=\frac{\gamma W_{\!\beta}}{r'}\,E(r'),\quad E^{\rho}=
\frac{\gamma\rho}{r'}\,E(r'),\quad B^{\varphi}=\beta E^{\rho},
\end{equation}
where
\begin{equation}
r'=\sqrt{\gamma^{2}W_{\!\beta}^2+\rho^{2}},\quad W_{\!\beta}=
Z+\beta T,
\end{equation}
and $E(r')$ is given by (\ref{Er}) with $r\to r'$. This field
exhibits some interesting features. It is of course
time--dependent since the source is in a uniform motion with
velocity $\beta$. Due to this motion a cylindrically symmetric
magnetic field, $B^{\varphi}$, arises. As $\beta$ increases
towards velocity of light, $E^Z$ becomes smaller and smaller,
whereas components $E^{\rho}$ and $B^{\varphi}$ approach the same
`$\delta$--profile'. The magnetic field becomes perpendicular to
the electric field and their magnitudes become equal. The field
concentrates on the null hyperplane $W_{\!\beta}\to 0$. It
resembles a plane wave, nevertheless, in contrast to the wave, the
invariant $F$ remains nonzero. Far from the charge the behavior of
the field is very nearly the same for all NLEs; as a consequence
of POC, it is of course very close to the field of the boosted
charge in the Maxwell theory. However, within a characteristic
radius of a particular NLE, where field equations and resulting
nonlinearities differ significantly, the specific features of the
NLE dominate. In Figure 1 the magnitude of electric field ${\rm
E}=\sqrt{(E^{\rho})^2+(E^{Z})^2}$ of a charge at rest and a charge
moving with velocity $\beta=0.9$ within the BI theory is
illustrated.  For comparison, see Figure 2 where the same
situation is displayed within the Maxwell theory. We see that the
behavior of the fields of static charges is reflected also in the
relativistic speeds. Although at the position of the charge $\rm
E$ diverges in the Maxwell theory, it remains finite in the BI
theory. In \cite{DaKu}, a number of other plots of boosted fields
are given for several specific NLEs. Another interesting effect
occurs in the very ultrarelativistic case. Now we proceed to
construct this limit.
 \begin{figure}
\begin{minipage}[]{8.5cm}
\hspace*{0.75cm}
    \rotatebox{-90}{
    \includegraphics[width=0.8\textwidth,height=0.2\textheight]{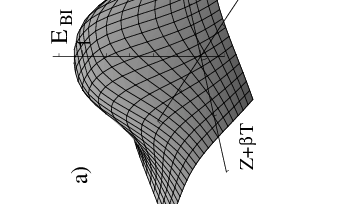}
     }
 \end{minipage}
 \begin{minipage}[]{8.5cm}
\hspace*{-1.12cm}
\rotatebox{-90}{
\includegraphics[width=0.8\textwidth,height=0.2\textheight]{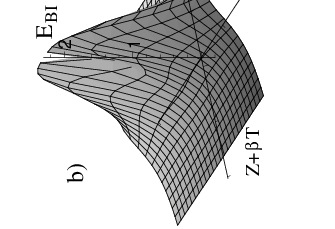}
}
 \end{minipage}
\vspace*{-1cm} \caption{The magnitude of electric field in
Born-Infeld theory. The graph a) displays ${\rm E}_{\rm BI}$ for a
charge at rest ($\beta=0$). The graph b) displays ${\rm E}_{\rm
BI}$ for a charge moving with velocity $\beta=0.9$. The units are
chosen so that $Q=1$, $r_0=1$.}
\end{figure}
\begin{figure}
 \begin{minipage}[]{8.5cm}
 \hspace*{0.75cm}
 \rotatebox{-90}{
\includegraphics[width=0.8\textwidth,height=0.2\textheight]{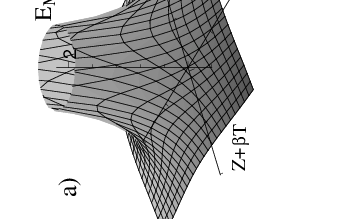}
}
\end{minipage}
 \begin{minipage}[]{8.5cm}
 \hspace*{-1.0cm}
 \rotatebox{-90}{
\includegraphics[width=0.8\textwidth,height=0.2\textheight]{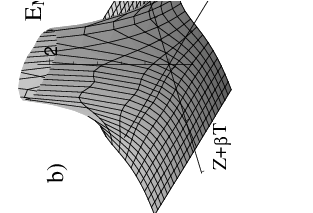}
}
\end{minipage}
\vspace*{-1cm} \caption{The magnitude of electric field in Maxwell
theory. The graph a) displays ${\rm E}_{\rm M}$ for a charge at
rest ($\beta=0$). The graph b) displays ${\rm E}_{\rm M}$ for a
charge moving with velocity $\beta=0.9$. The figure is plotted
with $Q=1$.}
\end{figure}

Let us first define function
$h(y)={\rho^2\!E(\sqrt{y^2+\rho^2})}/(2H(\rho){\sqrt{y^2+\rho^2}})$,
in which  function $H$ is chosen such that $\int\! h(y) dy=1$, so
that $h$ satisfies the assumptions of Lemma 1 in Section 2. Using
then both Lemma 1 and 2, and putting $\gamma=1/A$, we can derive
the component $E^{\rho}$ (resp. $B^{\varphi}$):
\begin{eqnarray}\label{}
\lim_{\beta\to 1-}\!\! E^{\rho}=\frac{2}{\rho}\,H(\rho)
\lim_{\beta\to
1-}\frac{1}{A}\,h\!\left(\frac{W_{\!\beta}}{A}\right)\
\stackrel{(L1, L2) }{=}\ \frac{2}{\rho}\,H(\rho)\delta(Z+T).
\end{eqnarray}
Regarding $E^{Z}$ one can prove that this component converges
pointwise to zero which implies the same convergence in the
distributional sense. (For the proof, see \cite{DaKu}.) Here we
proceed in an alternative way as follows:
\begin{eqnarray}
\lim_{\beta\to 1-}\!\! E^{Z}=
\lim_{\beta\to 1-}\frac{W_{\!\beta}}{\rho}\,E^{\rho}\ \stackrel{(L1, L2) }{=}\
\frac{2}{\rho^2}\,H(\rho)(Z+T)\delta(Z+T)=0,
\end{eqnarray}
where we used both Lemma 1 and 2, and standard relation $x\delta(x)=0$.
The results above enable us to formulate the following\\
{\bf Proposition.} {\it The electromagnetic field of a rapidly moving
charge constructed in a nonlinear electrodynamics with action
(\ref{S}) is in the ultrarelativistic limit, $\beta=1$, given by
\begin{equation}\label{ultr}
\mbox{\bf F}_{{\rm ultra}}=-E^{\rho}_{{\rm ultra}}dT\land d\rho-
B^{\varphi}_{{\rm ultra}} dZ\land d\rho,\quad
E^{\rho}_{{\rm ultra}}=B^{\varphi}_{{\rm ultra}}=\frac{2}{\rho}\,H(\rho)\delta(Z+T),
\end{equation}
where function $H$ is defined by
\begin{equation}\label{H}
H(\rho)=\frac{\rho^2}{2}\int_{-\infty}^{\infty}\!\!\!\!dy\,
\frac{E(\sqrt{y^2+\rho^2})}{\sqrt{y^2+\rho^2}},
\end{equation}
and function $E$ is implicitly determined by the relation
(\ref{Er})}.

The radial field $E$ is given only implicitly by equation
(\ref{Er}) and, in addition, even in cases in which it can be
determined explicitly it often has such a complicated form that
the integral (\ref{H}) can be evaluated only numerically. It is
thus useful to consider asymptotic series of this field for large
distances from the source. Regarding POC we assume the series to
be of the form
\begin{equation}\label{ser}
  E(r)=\frac{\alpha_{1}}{r^{2}}+\frac{\alpha_{2}}{r^{3}}+\dots=\!\!\sum\limits_{n=1,2\dots}
\!\!\frac{\alpha_{n}}{r^{n+1}},
\end{equation}
where $\alpha_1$ is equal to the charge $Q$ of the source and
coefficients $\alpha_2, \alpha_3\dots$ depend on the choice of a
particular NLE. Substituting this series into the integral
(\ref{H}) we obtain
\begin{equation}\label{Hser}
H(\rho)=\!\!\sum\limits_{k=1,2\dots}\!\left(\frac{\alpha_{2k-1}}{\rho^{2(k-1)}}
\frac{2^{2(k-1)}(k-1)!^{2}}{(2k-1)!}
+\frac{\alpha_{2k}}{\rho^{2k-1}}\frac{\pi(2k-1)!}{2^{2k}k!(k-1)!}\right)\!.
\end{equation}
The resulting field is then given by (\ref{ultr}) with $H(\rho)$
above. Specifically, for the Maxwell theory we have $H(\rho)=Q$,
and for the asymptotic expansion of the Born--Infeld field
(\ref{BI}) we get
\begin{eqnarray}\label{BIserie}
  H_{\rm BIser}(\rho)=Q\sum\limits_{k=0}^{\infty}\!\left(\begin{array}{c}
  \!\!\!-1/2\!\!\\\!\!\!k\!\!\end{array}\right)
\!\!\left(\frac{2r_{0}}{\rho}\!\right)^{4k}\!\!\!\frac{(2k)!^2}{(4k+1)!}=
  Q\!\left(\!1-\frac{4}{15}\frac{r_{0}^4}{\rho^{4}}+\frac{16}{105}
\frac{r_{0}^{8}}{\rho^{8}}\!\right)+O\!\left(\frac{r_0^{12}}{\rho^{12}}\right)\!.
\end{eqnarray}
In order to find the ultrarelativistic Born--Infeld field
everywhere we must evaluate the integral (\ref{H}) for the exact
expression $E_{\rm BI}$ given in eq. (\ref{BI}). After a simple
substitution this integral can be written in the form
\begin{equation}\label{BIultr}
H_{\rm BI}(\rho)=Q\tilde H_{\rm
BI}\left(\frac{\rho}{r_{0}}\right),\quad \tilde H_{\rm
BI}(x)=\frac{x^2}{2}\int\limits_{0}^{\infty}\!dt/{\sqrt{t(t+x^2)([t+x^2]^2+1)}}.
\end{equation}
We evaluated it numerically. The resulting field (\ref{ultr}) is,
up to the factor $\delta(Z+T)$, displayed in Figure 3. Here we
plot the $\rho$--components of ultrarelativistic fields obtained
within the Maxwell theory, $E^{\rho}_{\rm M}(\rm ultra)$, and in
the Born--Infeld theory using both the asymptotic expansion
(\ref{BIserie}), $E^{\rho}_{\rm BIser}(\rm ultra)$, and the
complete solution (\ref{BIultr}), $E^{\rho}_{\rm BI}(\rm ultra)$.
A striking difference between the Maxwell theory and the BI
electrodynamics is here clearly exhibited: the BI field becomes
zero at the origin whereas the corresponding Maxwell field
diverges as $E^{\rho}_{\rm M}\approx 1/\rho$. To find the behavior
of the BI field as $\rho\to 0$ analytically, we consider a
constant $\epsilon$, satisfying, $x^2\ll\epsilon\ll 1$. Then the
integral (\ref{BIultr}) can be approximated by
\begin{equation}
\tilde H_{\rm
BI}(x)\approx\frac{x^2}{2}\left(\int\limits_{0}^{\epsilon}\!\frac{dt}{\sqrt{t(t+x^2)}}
+\int\limits_{\epsilon}^{\infty}\!\frac{dt}{t\sqrt{t^2+1}}\right)
\approx -x^2\lg x + O(x^2).
\end{equation}
It is seen that the field approaches zero as $E^{\rho}_{\rm
BI}\approx -\rho\lg\rho$. Comparing the exact form of the BI field
with that obtained by employing the asymptotic expansion
(\ref{BIserie}) we observe that they both nearly coincide for
$\rho>r_0$ ($r_0=1$ in Figure 3) and approach the Maxwell field
asymptotically. (This can be shown also analytically \cite{DaKu}.)
From eqs. (\ref{ultr}), (\ref{H}) or (\ref{Hser}) examples of the
ultrarelativistic fields within other NLEs can be obtained.
\begin{figure}\label{3}
\rotatebox{-90}{
\includegraphics[width=0.74\textwidth,height=0.36\textheight]{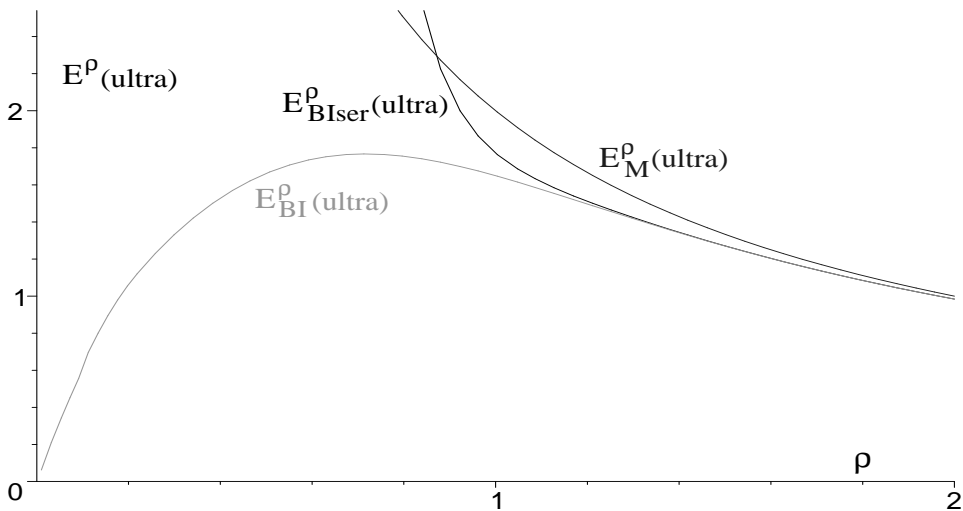}
} \vspace*{-4cm} \caption{The field of an ultrarelativistic charge
(BIon) in the BI theory. The $\rho$--components of the fields are
displayed up to the factor $\delta(Z-T)$ (cf. eq. (\ref{ultr})).
Three curves are plotted: one for the ultrarelativistic Maxwell
field $E^{\rho}_{\rm M}(\rm ultra)$, and two for the Born--Infeld
theory. $E_{\rm BIser}(\rm ultra)$ is constructed by using series
(\ref{BIserie}), $E_{\rm BI}(\rm ultra)$ is based on the full
exact solution (\ref{BI}), (\ref{ultr}), (\ref{H}). The units are
chosen so that $Q=1$, $r_{0}=1$. }
\end{figure}
\section{Concluding remarks}
The properties of the ultrarelativistic fields carry the
information about a theory of NLE from which they are derived.
Different NLEs lead to different types of ultrarelativistic
fields. For example, in the Born--Infeld theory the field remains
finite but nonvanishing for all relativistic velocities $\beta$,
but the field vanishes at the origin as $-\rho\lg\rho$ for the
very ultrarelativistic limit. In the Hoffmann--Infeld theory
\cite{Hoff} with the Lagrangian involving a logarithmic term the
electric field vanishes at the origin for all velocities. In all
cases satisfying POC at large distances the asymptotic behavior
$1/r^2$ of the fields of uniformly moving charges changes to the
`$1/\rho$--behavior' in the plane moving with charge with the
velocity of light in the ultrarelativistic limit.

The concept of the ultrarelativistic limit may be generalized to a
curved spacetime (see, e.g., \cite{Aich}, \cite{San},
\cite{Death}, \cite{Barabes}). Physically it appears plausible to
perform the limit while keeping the energy of the system fixed.
Then, however, one has to `renormalize' the fundamental constants,
e.g. mass $m\to m/\gamma$, charge $Q\to Q/\sqrt{\gamma}$; this
leads to a weak field regime. In this regime the Maxwell theory
appears to be fully satisfactory; the ultrarelativistic fields
within the Einstein--Maxwell theory were studied in \cite{San}.
Owing to the recent interest in BI and other NLEs, several authors
considered self--gravitating objects, in particular black holes,
within the Einstein--NLE theories (see, e.g., \cite{Gib},
\cite{ABG}). It would be interesting to study whether with such
spacetimes ultrarelativistic limits can meaningfully be formulated
while preserving the nonlinear character of the electromagnetic
theory.
\section{Acknowledgments}
We are grateful to Professors G. W. Gibbons and V. P. Frolov for
discussions. Financial support from the grand agency of the Czech
Republic GA\v CR--202/06/0041 and Prague Centre 
for Theoretical Astrophysics
is acknowledged.

\end{document}